\RequirePackage{fix-cm}
\documentclass[smallextended]{svjour3}       
\smartqed  
\usepackage{graphicx}
\usepackage{amsmath}
\usepackage{amssymb}
\usepackage{algorithm}
\usepackage{algpseudocode}
\usepackage{booktabs}

\begin{document}

\title{A Stochastic Approximation Approach for Foresighted Task Scheduling in Cloud Computing
}


\author{Seyedakbar Mostafavi        \and
        Vesal Hakami
}


\institute{Seyedakbar Mostafavi \at
              Department of Computer Engineering, Yazd University \\
              Tel.: +98-35-31232355\\
              \email{a.mostafavi@yazd.ac.ir}           
           \and
           Vesal Hakami \at
              School of Computer Engineering, Iran University of Science and Technology \\ 
}

\date{Received: date / Accepted: date}

\maketitle

\begin{abstract}
With the increasing and elastic demand for cloud resources, finding an optimal task scheduling mechanism become a challenge for cloud service providers. Due to the time-varying nature of resource demands in length and processing over time and dynamics and heterogeneity of cloud resources, existing myopic task scheduling solutions intended to maximize the performance of task scheduling are inefficient and sacrifice the long-time system performance in terms of resource utilization and response time. In this paper, we propose an optimal solution for performing foresighted task scheduling in a cloud environment. Since \emph{a-priori} knowledge from the dynamics in queue length of virtual machines is not known in run time, an online reinforcement learning approach is proposed for foresighted task allocation. The evaluation results show that our method not only reduce the response time and makespan of submitted tasks, but also increase the resource efficiency.  So in this thesis a scheduling method based on reinforcement learning is proposed. Adopting with environment conditions and responding to unsteady requests, reinforcement learning can cause a long-term increase in system's performance. The results show that this proposed method can not only reduce the response time and makespan but also increase resource efficiency as a minor goal.
\end{abstract}
\section{Introduction}\label{intro}
Cloud computing platforms are witnessing huge increase in user demands where a substantial part of computational tasks are offloaded to the cloud services due to its low Capex and Opex costs. Cloud service providers make it possible for end users to acquire and release computing and storage resource pools based on their changing needs over time. The number and diversity of tasks submitted by different applications to cloud services is increasing over time. Furthermore, the compute, storage and memory capacity of cloud service providers is heterogeneous and changing over the time. In this situation,  it is challenging to find an effective and efficient scheduling algorithm for appropriate task allocation to resources \cite{al2016performance}. The main problem here is how to assign user tasks to minimize makespan of a given set of tasks while providing quality of service (QoS) for all the tasks. Due to dynamics of clouds, how to find a successful task schedule with high utilization of cloud services is a substantial problem.

Task scheduling refers to mapping of tasks to cloud resources to improve system performance and meet quality of service requirements of applications including minimum makespan, minimum cost and high security and reliability. Achieving desirable quality of service is one of the challenging issues in task scheduling methods \cite{zhang2018dynamic}. Due to the exponential growth of solution space, the exhaustive search approach is not applicable to this problem. Most existing solutions for task assignment in cloud are myopic, meaning that these solutions are designed to minimize the instantaneous quality of service (i.e., the total makespan of submitted task for the next time interval). These solutions can be considered as minimizing makespan by repeated solving of an optimization problem. However, recent decisions for resource allocation impact the future system performance, which is not taken into consideration in these approaches. Optimal task allocation problem in cloud should be formulated as sequential decision making problem given unknown system dynamics. Therefore, the repeated optimization solutions are inferior to foresighted solutions that try to minimize the tasks makespan in long term.

The uncertainty associated with tasks submission process on the cloud systems is due to the random changes in tasks resource demands over time. By nature, the disposition of tasks over the virtual machines renders the system condition stochastic and time-varying. To account for the time-varying nature of task scheduling process, in this paper we propose a principled way to optimize the resource provisioning policy for cloud service providers through capturing these dynamics explicitly within a stochastic optimization framework with long-term objectives. We propose a model-free reinforcement learning algorithm which computes the optimal solution in the absence of the statistical information about the system, and instead, by relying only on the immediate feedback collected through real-time interactions with the environment. We design the learning agents' reward as a function of virtual machines' queue length which intuitively orients the system to the states with low VMs' queue length. Utilizing reinforcement learning approach in resource management makes it possible for scheduler to adapt itself with the dynamics in environment, considering time-varying nature of cloud resources and task length. Our scheduling algorithm takes into account task response time and makespan, trying to improve them considering the limited information about tasks and environment. Our main goal is to decrease response time and makespan and increase system utilization rate in the long-term. We seek an efficient policy that adaptively tunes the VMs' queue length as the control knob. Our objective is to minimize the average makespan and response time.  Our results show that in a cloud system with many tasks, our proposed method by applying the virtual machine queue length in reinforcement-learning improves response time in average by 49.5, 46, 43.9, 43.5 and 38.6 percent compared to random, Mixed, FIFO, greedy and Q-sch \cite{peng2015random} algorithms.

The rest of the paper is organized as follows. A comprehensive review over the state-the-art approaches for task scheduling in cloud systems is presented in Section \ref{related}. The system model and the problem formulation of reinforcement learning algorithm is presented in Sections 3 and 4, respectively. In Section 5, we elaborates on the proposed method, designing our task scheduling algorithm. The evaluation results are presented in Section 6 and the paper is concluded in Section 6. 

\section{Related Work} \label{related}
Since the computational task offloading over cloud was popular, a large body of literature on the cloud computing has been devoted to the problem of optimal task scheduling on the cloud platforms. Based on the availability of information in the cloud platform, the literature on the problem of task scheduling in cloud environments can be classified into three categories: offline optimization, online optimization and model-free schemes. In offline optimization, it is assumed that the exact information about the system configuration (host and VM settings), user task distribution and QoS parameters is available in the cloud platform. On the other hand, an online optimization framework utilizes statistical information about the system data, parameters and computation tasks in the system. Due to the hardness of finding optimal policy for task scheduling in the form of offline optimization formulation, much research efforts have been done in the recent years to find the near-optimal solutions through proposing dynamic online algorithms. 

The most state-of-art solutions for task scheduling are based on the heuristic and meta-heuristic techniques including ant colony\cite{azad2019fuzzy}, PSO \cite{ebadifard2018pso}, genetic \cite{keshanchi2017improved} and fuzzy \cite {azad2019fuzzy},\cite{guo2015workflow} algorithms. Although these algorithms are effective in finding approximate solutions, they could not capture the long term evolution of system performance. Model-free approaches mostly utilize learning-theoretic schemes, e.g. reinforcement learning techniques. The authors in \cite{peng2015random} present a model for optimal task scheduling based on reinforcement learning method by dividing the problem into sub models. In this method, task scheduler exploits a Q-learning strategy to make proper task distribution over virtual machines which results in response time improvement. Although this solution decreases the task response time, it does not consider the queue length of virtual machines which causes longer total makespan for all the submitted tasks. In order to support program scalability in cloud, a reinforcement learning-based method was proposed in \cite{barrett2013applying} for optimal resource allocation. This approach leads the agent to learn optimal policies for task allocation, but it does not consider the diversity in users' requests. Also, this solutions suffers from the curse of dimensionality that makes it unsuitable for real world problems. In \cite{yang2011utility}, a task scheduling solution considering the economic utility of agents is introduced which models the resource scheduling problem by considering failure rate and recovery rate. The results of this paper reveal that this scheduling algorithm is efficient and results in optimal system efficiency. This article investigates only system efficiency and ignores other quality of service parameters such as response time and makespan. The problem of task scheduling has been also addressed in hybrid cloud-fog environments \cite{pham2016towards}\cite{mostafavi2019fog}. Optimization of energy consumption is another challenging issue in task scheduling. The authors in \cite{hussin2011efficient} propose an energy management framework in cloud based on reinforcement learning methods which takes into account the heterogeneity in cloud resources and users' requests. In paper \cite{hussin2015improving}, a scheduling method based on reinforcement learning is proposed with the purpose of providing reliable service and improving system performance with low computational complexity. In \cite{cheng2018drl}, DRL-Cloud is proposed as a deep reinforcement-learning based approach for energy cost minimization for cloud service providers. In this approach, a joint resource provisioning and task scheduling algorithm is designed which generates the long-term decisions for resource allocation. Other similar deep learning based solutions for cloud task scheduling has also been proposed in \cite{wei2018drl}, \cite{lin2018deep} and \cite{li2019deepjs}. The authors in \cite{chen2019reinforcement} deal with the problem of multi project scheduling in cloud manufacturing systems. Intelligent task scheduling strategies has been proposed in \cite{xue2019intelligent} and \cite{liu2018reinforcement} for cloud robots which divides the complex reinforcement learning problem into sub-problems to find the task scheduling policy. This approach, yet converges quickly to the policies for sub-problems, fails to achieve the optimal long-term solution. A general reinforcement learning algorithm for scheduling problems in heterogeneous distributed systems has been proposed in \cite{orhean2018new} which models the task dependencies on the grid with acyclic graphs to find the optimal solution through implementing machine learning box. Wang \textit{et. al} formulate the problem of multi-workflow scheduling over infrastructure-as-a-service clouds as a Markovian game and propose a Q-learning algorithm to find the optimal policy. The authors in \cite{duggan2017reinforcement}, \cite{cui2018cloud}, \cite{wei2017reinforcement}, \cite{zhong2019multi} and \cite{nascimento2019reinforcement} have put some efforts to solve the problem of service composition and workflow scheduling through learning-based approaches. In  \cite{balla2018reliability} an action-selection method is proposed to preserve reliability in cloud systems through stable resource provisioning. Sahoo \textit{et. al} consider the time sensitivity of tasks in the cloud, proposing a learning-automata algorithm based on Lyapunov optimization to find the near optimal solution for bi-objective minimization problem of energy consumption and makespan. 

There is a body of works in P2P-cloud systems \cite{mostafaviijcs} \cite{mostafavi2016game} \cite{mostafavi2017stochastic} which apply the model-free reinforcement learning techniques for resource allocation in P2P-cloud platforms. In \cite{cui2017reinforcement} a cloud architecture is proposed for optimal job scheduling under SLA constraints.  A reinforcement-learning-based cloud auto-test systems is designed in paper \cite{zhao2017dynamic} which aims to maximize the long-term benefits of the system. The reinforcement learning also has been utilized as a main theoretical framework to model the problem of resource provisioning in the cloud \cite{salahuddin2016reinforcement}\cite{liu2017hierarchical}\cite{zhang2017intelligent}. Its inherent characteristics including self-adaptive decision making and long-term utility maximization fits well with the time-varying and dynamic conditions of cloud environments.

\section{System Model}\label{system}
In this section, we describe the settings of a task scheduling mechanism in a cloud service provider. The cloud broker as task scheduler plays the central role of an agent which interacts with the environment. The task scheduler should make optimal decisions to achieve the time constraints of tasks in long term. The general broker model comprises from three main parts: task transmission, task allocation and task execution. In the transmission phase, the scheduler assigns users' tasks from the global system queue to one of the queues of virtual machines (VMs). The global queue accepts users' request over time and queues them in a first-come-first-serve manner and chooses them for processing. Every VM has a buffer queue with a stable capacity. Task transmitter knows \emph{a-priori} the buffer capacity of VMs for allocating the tasks. If the number of VMs equals to K and the buffer capacity of each VM equals to N, then the occupied capacity of system S will be determined as follows:
\begin{eqnarray}0\leq\overset k{\underset{i=0}{\sum S_i}}\leq K\times N 
\end{eqnarray}

\begin{figure}[h]
\centerline{\includegraphics[width=0.8\textwidth]{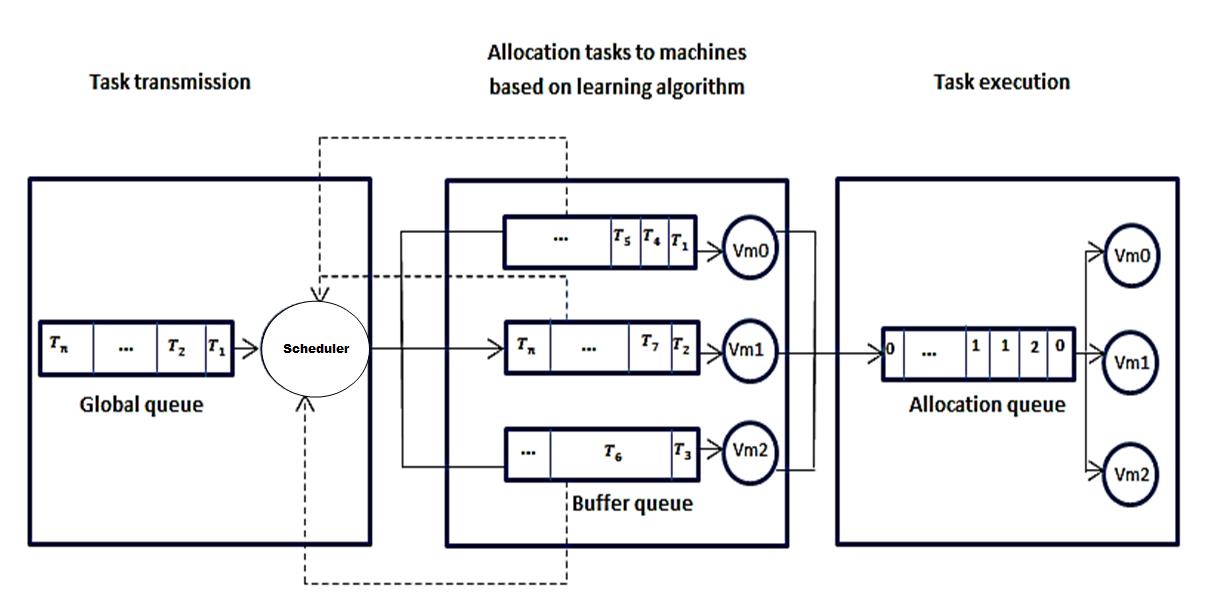}}
\caption{The system model}\label{fig1}
\end{figure}

Referring to Figure (\ref{fig1}), the scheduler first should choose from the global queue the tasks should to be assigned to VMs. For each chosen task, the scheduler observes the state of the environment and selects one of the virtual machines for task mapping based on the reinforcement learning algorithm. Mapping of each task to a VM results in a reward for the chosen action. Then, the scheduler stores the pairs of (action,reward) in the memory for the future uses. After task allocation to virtual machines, the buffer capacity of the VMs will be updated accordingly. By arriving next request, the scheduler meets a new state of the environment and decides based on the state-action table until all tasks are allocated to VMs. Finally, after a proper mapping of tasks, they are sent to the next queue (allocation queue) to be executed. Allocation queue is an array which its indices are task numbers and the values are the allocated VMs over which the task should be executed. This array represents which tasks have been assigned to which virtual machine, therefore it should be configured so that requests bear minimum waiting time in the buffer queue and it results to minimum response time. We assume that system operates in discrete time, i.e. the time is divided into normalized slots $n\in \{0,1,2,...\}$. In each time slot, the broker receives a random number of tasks from the users. The state space of number of received tasks is represented as a discrete and finite space $\textit{D}=\{0,1,2,...,D\}$. The number of tasks arrived to the cloud at time slot \textit{n} is denoted by $d_n \in \textit{D}$. 

\textbf{Assumption 1}. The dynamics of task arrival into the cloud system follows a Markov chain. It worth to note that a simpler assumption for task arrival would be the case where $\{d_n\}_{n\in \mathbb{N}}$ is i.i.d.

\section{The Proposed Stochastic Optimization-based Solution}\label{sec:proposed}
Given the system model in Section \ref{system}, we aim to design a task assignment policy that minimizes the expected cumulative response time of tasks. The devised control policy should be adaptive to the dynamic states of the system, i.e., to the time-varying volume of tasks. In particular, adaptability to volume of tasks is needed to make policy aware from delay under the conditions of unsaturated traffic and finite-length queue at the virtual machines. Considering the stochastic dynamics of cloud system, choosing myopic actions about the queue length of VMs based only on the immediate observations cannot lead the system to the optimal policy. In fact, the task scheduling problem is a sequential decision making problem in the sense that it would be rational for cloud broker that sometimes make instantaneous suboptimal decisions is some time periods for inducing the system states into the desirable states. 

\subsection{MDP-based formulation}
To find the optimal policy in sequential decision problem, we systematically formulate the task scheduling problem into the stochastic optimization framework of Markov Decision Process (MDP) \cite{sutton2018reinforcement}, given the Markovian nature of our settings. Formally, MDP associated with task scheduling problem is formulated as a tuple $C=(S, A, P, R)$ where $S, A, P$ and $R$ denotes respectively the set of states, the set of control actions, transition probabilities, and reward function. Specifically, the system state space is a discrete and finite space as the Cartesian product of VMs' occupied buffer and total task length, i.e., $\textit{S} = \textit{B} \times \textit{L}$ and $s_t=\{b_n,L_n\} \in \textit{S}$ represents the system state in time slot $\textit{t}$. 

In the following, we describe the control actions, state transition law, the reward function, as well as the formulation of the stochastic optimization problem.   
\begin{itemize}
\item State space $S = \{B_1,\dots,B_k,L_1,\dots,L_k\}$: the state in task scheduling problem is defined as a tuple comprised from VMs' occupied buffer and the total length of assigned tasks to VMs where $S_k$ represents the occupied buffer of virtual machine and $T_k$ represents the length of allocated tasks to $k $th virtual machine.
\item Action space $A = \{a \in \mathbb{N} | 0 \leq a \leq M \}$: an action in MDP is the virtual machine which is selected by the broker for task excecution, which is bounded to $M$, maximum number of VMs in the experiments. 
\item Probability distribution $P : S \times A \times S \rightarrow [0,1]$: Following the execution of action $a_t$ in state $s_t$, the system state transitions probabilistically to the next state $s_{t+1}$. The transition probability $P(s_{t+1}|s_t,a_t)$ is determined according to the following equation:
\begin{eqnarray}
P(s_{t+1}|s_t,a_t) = P(B_{t+1}|B_t,L_t) \cdot P(L_{t+1}|L_t)
\end{eqnarray}
\item Reward function $R: S\times A \times \mathbb{R}\rightarrow [0,1]$: A value function which measures the goodness of a chosen action $a$ when we are in the state $s$. A discount factor $0 \leq \gamma \leq 1$ is considered to compute the expected reward of system in infinite iterations. The expected reward function for a pair $(s,a)$ is defined as follows:
\begin{eqnarray} Q(s,a) = E\{\sum_{k=0}^{\infty}{\gamma^k r_{t+k+1}|s_t=s, a_t=a}\} 
\label {eq1}\end{eqnarray}

The value function defined in (\ref{eq1}) is often referred to as $Q$-value function. The instant reward achieved by each agent for the taken action $a$ in the state $s$ is defined as follows:
\begin{eqnarray}
R(s,a) = \begin{cases}
		1 &  \text{if task with length $L_k$ is assigned to a VM with$ min (B_k)$} \\
		-1 & \text{if task with length $L_k$ is assigned to a VM with $max (L_k)$} \\
		0 & otherwise
	\end{cases}
\end{eqnarray}
where $L_k$ is the length of tasks assigned to the virtual machine $k$. If the newly arrived task is assigned to a VM with minimum queue length, the reward will be 1. On the other hand, if task was assigned to a VM with maximum task execution length, the reward will be -1. In other cases, the agent will receive a reward equals to 0. 

\item Optimization objective function: The objective function in our MDP formulation is to maximize the long-term average discounted reward $R(s)$. When the $P$ and $R$ functions  are determined \emph{a-priori}, the classical Dynamic Programming algorithms such as value iteration \cite{kaelbling1996reinforcement} can be utilized to find the optimal policy. In value iteration method, the expected total reward of system over a finite horizon is computed as follows:
\begin{eqnarray}
V^*(s)=\underset{a}{\mathrm{max}} {\Bigl [R(s,a)+\gamma \int_{s'\in S}{P(s,a,s')V^*(s')}\Bigr]}
\end{eqnarray}
This methods applies successive approximations until a predetermined error bound $\epsilon$ is reached. The classical dynamic programming method \cite{bertsekas1995dynamic} computes the policies in an offline mode. The cloud broker seeks to learn the optimal policy $pi^*$ to adjust task assignment strategy, so that for all $s\in S$, the optimal policy is computed by:
\begin{eqnarray}\label{objective}
\pi^*(s)=\underset{\pi}{\mathrm{argmax}} {\bar{R}^{\pi}(s)} \overset{\Delta}{=} \mathbb{E}^{\pi} \Bigl [ \sum_{t=1}^{\infty}{\gamma^t R(s_t,a_{s_t})|s_1=s} \Bigr]
\end{eqnarray}
where $\bar{R}^{\pi}(s)$ represents the average discounted sum, in which the discount factor $0\leq\gamma\leq1$ represents the importance of future rewards proportional to the current reward. The smaller values for $\gamma$ gives less importance to the future rewards, thus rendering the broker's behavior toward more myopic policy. 
\end{itemize}
 
\subsection{Learning the optimal policy}
In the online decision-making settings where the reliable estimations about the underlying stochastic processes are not available, we cannot apply a model-based scheme (e.g. standard value iteration\cite{bertsekas1995dynamic}) to solve the problem in \ref{objective}, because such a scheme relies on the information about the system's transitions. Instead, in this section we propose a model-free learning algorithm to find the optimal policy in the absence of statistical information of the system and merely based on the immediate feedback acquired through  real-time interactions with the system. 
 
\begin{figure}[ht]
\centerline{\includegraphics[width=0.8\textwidth]{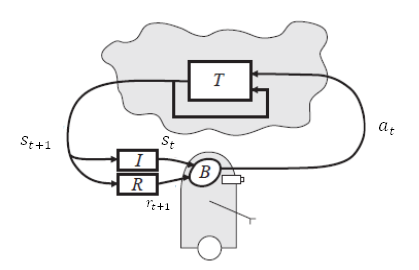}}
\caption{The reinforcement learning mechanism}\label{fig2}
\end{figure}

Specifically, we implement a learning algorithm in the broker which iteratively estimate $\pi^*$ for solving the problem \ref{objective}. The problem \ref{objective} is solved using standard Q-learning \cite{sutton2018reinforcement} which gradually estimates the optimal policy $\pi$. In Q-learning, the learning agent repeatedly takes an action $a$ given an observed state $s$, receiving the reward $R(s)$ and it is transitioned to a new state $s'$ (see Figure \ref{fig2}). For completeness, we first present standard Bellman equations \cite{peng1992stochastic} associated with our problem, and then elaborate on the Q-learning algorithm. The so-called $Q$ function is denoted by $Q^*(s,a)$ which is defined as the sum of immediate reward $l(s,a)$, obtained from by taking action $a$ in state $s$ plus the long-term reward obtained from following the optimal policy $\pi^*$; i.e., 
\begin{eqnarray}\label{bellman}
Q^*(s,a) = L(s,a) + \gamma \cdot \sum_{s'\in S}{\mathbb{P}(s'|s,a)\bar{L}^{\pi^*}(s')}
\end{eqnarray}
where
\begin{eqnarray}
\bar{L}^{\pi^*}(s') = \underset{a\in \mathbb{A}(s')}{\mathrm{max}} {Q^*(s,a)}, \forall s \in S
\end{eqnarray}

Then, the optimal policy $\pi^*(s)$ is obtained as:
\begin{eqnarray}
\pi^*(s) = \underset{a\in \mathbb{A}(s)}{\mathrm{min}} {Q^*(s,a)}, \forall s \in S
\end{eqnarray}

Unlike a model-based learning algorithm which requires information of $\mathbb{P}$ for solving the Bellman equations in \ref{bellman}, Q-learning does not need such knowledge through repeated estimation of $Q^*(s,a)$ for all pairs $(s,a)$. In this algorithm, the estimate table $\hat{Q}(s,a)$ is initialized with zero or arbitrary values for all pairs $(s,a)$. In each iteration, the broker observes the current state $s$ and chooses an action $a$ from action space. After implementing the selected action, the achieved immediate reward $l(s,a)$ is computed and then, the estimate $\hat{Q}(s,a)$ for current pair $(s,a)$ is updated according to the following equation:
\begin{eqnarray} \label{q-equation}
\hat{Q}_t (s,a) \leftarrow (1-\beta_t)\hat{Q}_{t-1}(s,a) + \beta_t \Bigl[l(s,a) + \gamma \cdot \underset{a' \in \mathbb{A}(s')}{\mathrm{min}} {\hat{Q}_{t-1}(s',a')}  \Bigr]
\end{eqnarray}

In \ref{q-equation}, $\hat{Q}_t (s,a)$ represents the $t$-th estimation of $Q^*(s,a)$ and $\beta{t}$ represents the learning rate of Q-learning algorithm, which is computed for each pair $(s,a)$ according to \ref{beta}:

\begin{eqnarray} \label{beta}
\beta_t = \frac{1}{1+ (visit_t(s,a))^{0.65}}
\end{eqnarray}

where $visit_t(s,a)$ represents the number of times the pair $(s,a)$ has been observed up to iteration $t$. We first assume that the virtual machines are homogeneous in respect of RAM, CPU and buffer size capacities. For example, for three VMs, the initial state met by the scheduler will be represented by $VM_3 = (0,0,0,0,0,0)$ that three first parameters representing the occupied buffer and the next ones shows sum of the assigned task length for $VM_3$, respectively. The scheduler in each time step, after observing the system state, chooses an action $a$ from the action set $A(s) = \{1,2,3\}$ based on a $\epsilon$-greedy policy.

The necessary condition for convergence guarantee of described Q-learning algorithm is to choose actions at each iteration in a $\epsilon$-greedy fashion \cite{djonin2007q}. $\epsilon$ is a small probability with which the learning agent explores the unknown environment. Specifically, although the broker needs generally take greedy actions according to the optimal policy being estimated, it sporadically should take random actions to further explore the alternative actions. These behavior guarantee that the agent does not get biased towards actions with misleading good Q values (equation \ref{epsilon}):

\begin{eqnarray} \label{epsilon}
			\pi (s) = \begin{cases}
			\text{random action from } \mathbb{A}(s), & \text{if }  \xi < \epsilon \\
			\underset{a}{\mathrm{argmax}} {Q(s,a)}, & \text{otherwise}
			\end{cases}
\end{eqnarray} 
where $0 \leq \xi \leq 1$ is a uniform random number drawn at each time step. 

Using such $\epsilon$-greedy action selection policy, the described Q-learning algorithm is guaranteed to converge to $\pi^*(s)$. 

\subsection{State Space Reduction}
In reinforcement learning algorithms, dealing with the exponential growth of state space is a challenging problem. This is known as curse of dimensionality. Due to huge state space in large scale cloud systems with many virtual machines, learning rate is slow and the algorithm may not converge to the optimal solution. In this paper, we propose a novel fuzzy-based state reduction solution which discretizes the task execution length to reduce the state space dimension. To this end, we approximate the duration of task $i$, $D_{L_i}$, over a specified range for task length as follows:
\begin{eqnarray}
Length_i = \Bigl [ L_i/ Range\Bigr ]
\end{eqnarray}
This fuzzy approximation results in huge reduction in the state space and improves the algorithm performance, as shown in the Evaluation section.
\subsection{Adaptive $\epsilon$-greedy Exploration}
Balancing the ratio of exploration and exploitation is an important challenge in reinforcement learning which greatly affects the learning time \cite{abbeel2005exploration}. The agent can choose to explore the environment and try new choices to find a better policy for the future, or exploit the already tested actions. A pure strategy for exploration or exploitation can not result in best policies. So, the agent should select from a balanced combination of exploration and exploitation to achieve the best results. 

In this paper, we address the dilemma of exploration and exploitation with an adaptive approach which controls the exploration. In this approach, the agent is biased to explore more in the initial steps of algorithms to find the better candidate. The achieved experience from the previous steps are exploited more in the final steps to focus on the best available solutions. For this purpose, the value of $\epsilon$ is initialized with a random value in $[0,1]$. Then, the value of this variable is updated considering the remaining steps to convergence as follows:
\begin{eqnarray}
\epsilon = \epsilon - \dfrac{\epsilon}{total cycles}*cycle number
\end{eqnarray}
where total cycles represent the maximum number of cycles for convergence and cycle number shows the current cycle. In this way, the $\epsilon$ will gradually converge to zero and the algorithm will choose the action more greedily. 

\subsection{QL-Scheduling Algorithm}

Following the formulation presented in Section \ref{sec:proposed}, we describe the Q-Learning-Scheduling algorithm. The action selection phase follows a $\epsilon$-greedy solution in which the selected action in each time step is a random action with probability $\epsilon$, or the optimal action with probability $1-\epsilon$. The pseudo-code of task scheduling algorithm is represented in Algorithm \ref{table:algorithm}.

\begin{algorithm}
\caption{Pseudocode of the proposed algorithm}\label{table:algorithm}
\begin{algorithmic}[1]
	\State Q-table is initialized to 0
	\State The \textit{best action} and \textit{repeater} parameters are initialized to 0
	\Repeat { for each episode}
	\State Initialize $s_t$
	\Repeat { for each step of episode}
	\State $a_t$ = actionSelection($s_t$,Q) using $\epsilon$-greedy algorithm
	\State Take action $a_t$ and observe reward $r_t$
	\State $Q(s_t,a_t) = Q(s_t,a_t)+\alpha [r_{t+1}+\gamma \underset{a}{\mathrm{max}} {Q(s_{t+1},a_{t+1}-Q(s_t,a_t))}]$
	\State Update state as $s_t = s_{t+1}$
	\Until { $s_t$ is a terminal state}
	\State repeater = repeater + 1
	\State Investigate the best action and repeater 
	\Until { best action does not change or $repeater > \theta$ }
\end{algorithmic}
\end{algorithm}

The details of algorithms are explained as follows:
\begin{enumerate}
\item The Q-table is initialized for all the state-action pairs to zeros.
\item A counter, \textit{repeater} is defined as a threshold for the maximum number of state visits and an array, \textit{best action}, is defined to determine the best chosen action in each time step. These variables are used to analyze the convergence of algorithm.
\item In steps 3-5, two inner loops are defined to iterate over episodes and time steps in each episode.
\item In step 6, the best action $a_t$ is chosen after observing the current state $s_t$ based on a $\epsilon$-greedy policy.
\item The chosen action $a_t$ is taken, and the reward value $r$ and the next state $s_{t+1}$ is determined.
\item The Q-table for the ($s_t$, $a_t$) pair is updated accordingly.
\item The new state $s_{t+1}$ is considered as the current state of the system.
\item The steps 6-9 are repeated until the terminal state is visited.
\item The \textit{repeater} variable is increased.
\item The best action and repeater variables are considered to determine the convergence of algorithm.
\item The algorithm is finished when the \textit{best action} does not change or the \textit{repeater} is exceeded from the repeater threshold. 
\end{enumerate}

\subsection{Algorithm Complexity}
Our proposed QL-Scheduling algorithm for task-scheduling (Algorithm 1) is a particularly lightweight algorithm: the algorithm’s update rules are written in terms of efficient recursive formulae, which lead to negligible complexity. In fact, at each time step, the algorithm needs just a few standard algebraic operations, along with one random number generation to calculate the next action (based on the $\epsilon$-greedy policy). Despite its minimal per iteration complexity, however, similar to all reinforcement learning algorithms with stochastic convergence guarantees, the rate of convergence is sub-linear at best. 

\begin{remark}
In the proposed mechanism, we have devised no specific fault tolerance mechanism to detect and recover from a fault that is happening or has already happened in either software or hardware in the system in which the software is running. However, despite the lack of a mechanism for handling machine failure, we argue that the job completion time of our algorithm would not be very large when server failure occurs. In our proposed algorithm, in cases where some task fails after starting, it would be requeued to be restarted shortly after failing. Also, given that the operational state of the system is comprised of the VMs' occupied buffer as well as the total length of assigned tasks to VMs, the tasks submitted to the failed machine would soon be rescheduled to another low load server to decrease the makespan. It has been shown in the section \ref{evaluation} that the average makespan resultant from our approach is fairly robust against failures.
\end{remark}

\begin{remark}
The “repeater” parameter reflects the total “iteration budget” available to the algorithm. The higher the budget, the higher would be the number of samples that can be collected by the learning algorithm to update its Q-value estimates as well as its control policy. Obviously, with higher sample complexity, the quality of the converged solution would increase accordingly. 
It is up to the specific application context and to the administration of the cloud data center how to limit the maximum number of iterations when using our algorithm for task scheduling. There can even be no constraint on the repeater, as typically all reinforcement learning procedures run in the form of an infinite loop, constantly learning and adapting to variations that might occur in the operational environment. However, in certain cases with limited computational or energy resources, “repeater” can be capped by a threshold. The impact of setting lower values for “repeater” can be readily observed from Figure \ref{fig:convergence} for instance, where in earlier time steps, the average response time is much higher compared to later iterations. 
\end{remark}

\section{Evaluation Results}\label{evaluation}
In this section, we perform exhaustive experiments over CloudSim simulator to evaluate the performance of the proposed algorithm. For this purpose, the convergence of algorithm is analyzed first. Then, we design two complementary scenarios to compare the response time, waiting time, makespan and utilization rate of the proposed Q-Learning-Scheduling algorithm with the state of the art solutions. Furthermore, the impact of buffer size over the proposed solution are discussed. We have implemented the following algorithms for the purpose of comparison:
\begin{itemize}
\item The proposed algorithm Q-learning which utilizes a reinforcement-learning policy to optimize the task scheduling based on a $\epsilon$-greedy policy.
\item Random algorithm \cite{kowsigan2019efficient} which randomly distribute the tasks over the virtual machines. In this method, if there is not buffer capacity in the virtual machines, the tasks are rescheduled and they are assigned to the virtual machines with enough buffer size.
\item First-In-First-Out (FIFO) scheduling mechanism \cite{FIFOisard2007dryad} in which the tasks are assigned to the virtual machines based on their order of arrival. This method performs well when the tasks are arrived in a ascending order in terms of their size and has poor performance if the tasks are arrived in a descending order.
\item Mixed scheduling mechanism \cite{peng2015random} in which the tasks are first assigned to the virtual machines in a random fashion, but if there is unused buffer size in virtual machines, the tasks are rescheduled to be assigned to the virtual machine with maximum buffer capacity. 
\item The greedy scheduling method \cite{dong2015greedy} where the tasks are assigned to the best virtual machine in terms of the unused buffer capacity. 
\item The Q-sch method proposed in \cite{peng2015random} which utilizes the q-learning method for task assignment to the virtual machines. In this method, the state is composed from the unused capacity of virtual machines. It is assumed that the buffer of virtual machines is empty in the start of algorithm and the buffer is decreased when a task is assigned to the virtual machine.  
\end{itemize}
\subsection{Convergence Analysis}
\begin{figure}
\centerline{\includegraphics[width=0.8\textwidth]{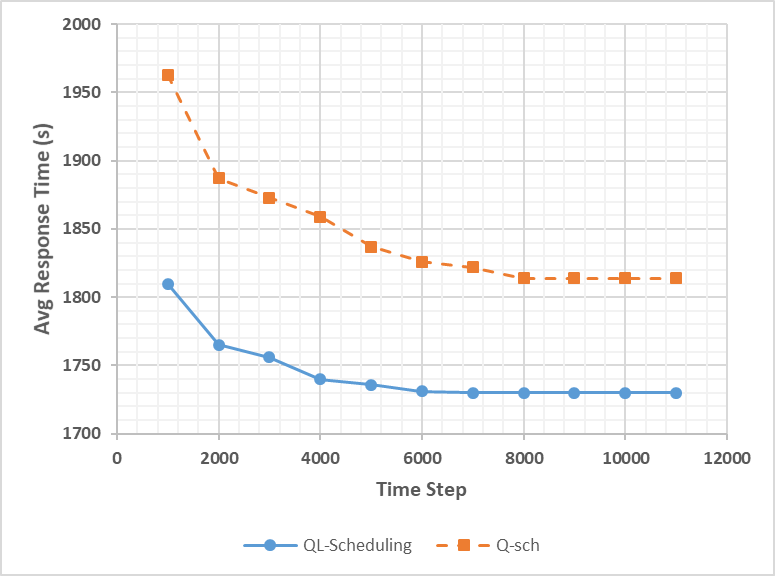}}
\caption{The convergence analysis of QL-Scheduling algorithm}\label{fig:convergence}
\end{figure}

In the proposed QL-Scheduling algorithm, the scheduler learns from the previous task assignments the optimal policy. This process is repeated until the algorithm converges to a stable action selection policy. The algorithm will finish when the final states are visited for enough times and a unique action selection policy is achieved. For convergence analysis, we have evaluated the waiting time for 50 tasks over different time steps. As shown in Figure (\ref{fig:convergence}), the waiting time values are fluctuating over time steps between 1000 and 4000 time steps. However, the algorithm converges to the optimal policy from time steps 5000 onward with average waiting time of 1730. In our proposed algorithm, two metrics are considered to check the convergence of algorithm. In addition to maximum number of time steps for visiting the states, we check the chosen action in the current time step to see if it has changed from the previous time step. It is also observed from the Figure \ref{fig:convergence} where QL-Scheduling converges faster than Q-sch algorithm in terms of number of time steps. 

\subsection{Evaluation Scenarios}
In this paper, two lightweight load and heavy load scenarios are described to assess the performance of the proposed algorithm. In the first scenario, task arrival rate is changed from 10 to 20 tasks. The other specifications of this scenario is represented in Table \ref{table:firstscenario}. In the second scenario, much more tasks (from 20 to 100) are arrived to the system and the length of tasks has much more variations. The parameters of the second scenario is show in Table \ref{table:secondscenario}. In simulations, two hosts are setup containing different number of virtual machines with specified buffer sizes. It is assumed that the virtual machines are homogeneous in terms of system specifications and the tasks are injected to the system from a real-world trace. The tasks are randomly chosen from the trace file so as the tasks are different in terms of length. For fair comparison, each simulation scenario is repeated several times and the average values of results are shown in the following subsections. 

\begin{center}
\begin{table}
\centering
\caption{Parameters of first scenario.\label{table:firstscenario}}%
\begin{tabular*}{320pt}{@{\extracolsep\fill}lcccc@{\extracolsep\fill}}
\toprule
\textbf{Parameters} & \textbf{Values} \\
\midrule
Task length & 5000-200000 million instructions  \\
Total number of tasks & 20   \\
Total number of VMs & 3  \\
VM frequency & 1000 MIPS \\
VM memory (RAM) & 1740 MB \\
VM bandwidth & 1000 MBps \\
VM buffer size & 5-15 tasks \\
Number of PEs & 1 \\
Number of data centers & 1 \\
Number of hosts & 1	\\
\bottomrule
\end{tabular*}
\end{table}
\end{center}
\begin{center}
\begin{table}
\centering
\caption{Parameters of second scenario.\label{table:secondscenario}}%
\begin{tabular*}{320pt}{@{\extracolsep\fill}lcccc@{\extracolsep\fill}}
\toprule
\textbf{Parameters} & \textbf{Values} \\
\midrule
Task length & 100-400000 million instructions  \\
Total number of tasks & 100   \\
Total number of VMs & 3  \\
VM frequency & 1000 MIPS \\
VM memory (RAM) & 1740 MB \\
VM bandwidth & 1000 MBps \\
VM buffer size & 5-50 tasks \\
Number of PEs & 5 \\
Number of data centers & 1 \\
Number of hosts & 2	\\
\bottomrule
\end{tabular*}
\end{table}
\end{center}

\subsection {Evaluation Metrics}
Several factors are considered in the literature for evaluating the performance of task scheduling algorithms. In this paper, we evaluate the most important factors including response time, waiting time, makespan, utilization rate and load distribution. Response time is defined as the duration from task assignment to VM up to task completion. Waiting time represents the stall time in the VMs buffer until task execution is started. Makespan is defined as the total time from the arrival of first task last task is executed successfully in the system. The average values of response time, waiting time and makespan are calculated in the simulations based on the following equations, respectively:
\begin{eqnarray}
\text{Average response time} = \frac{1}{N} \sum_{i\in Tasks}{(F_{T_i}-S_{T_i})} \\
\text{Average waiting time} = \frac{1}{N} \sum_{i\in Tasks}{(F_{T_i}-S_{T_i}-E_{T_i})} \\
\text{Average Makespan} = \underset{i\in Tasks}{\mathrm{max}} {M_{T_i}}
\end{eqnarray}  
where $F_{T_i}$ represents the task finish time, $S_{T_i}$ represents the task start time, $E_{T_i}$ shows the task execution time, and $M_{T_i}$ represents the finish time of all tasks. The total number of tasks are represented by $N$.

\subsection{Average Makespan}
\begin{figure}
\centerline{\includegraphics[width=0.8\textwidth]{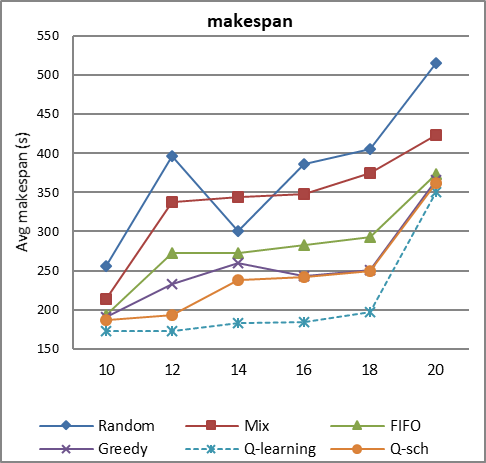}}
\caption{The average makespan in 1\textsuperscript{st} scenario}\label{fig:makespan1}
\end{figure}

In cloud computing platforms, the average makespan is an important factor which has great effects on the quality of service experienced by the users. lower value of average makespan means that there is less delay for task execution completion and more tasks will be completed in a specified time interval. In our proposed method, we provide the system with an optimal task allocation policy in which the short length tasks do not wait for longer tasks in the queue and therefore, the average makespan of systems is substantially reduced. As shown in Figure (\ref{fig:makespan1}), the QL-Scheduling algorithm achieves lower makespan in the experiments of first scenario compared to the competing algorithms. The proposed algorithm works well even in the experiments where the tasks are arrived in ascending order. It outperforms the Q-sch algorithm in the first scenario with 16 tasks with 23.56\% improvement. Improvement  values of QL-Scheduling comparing to the other algorithms are represented in Table \ref{makespan1}.

\begin{center}

\begin{table}
\centering
\caption{Improving makespan of the proposed algorithm (1\textsuperscript{st} scenario).\label{makespan1}}%
\begin{tabular*}{320pt}{@{\extracolsep\fill}lccccccc@{\extracolsep\fill}}
\toprule
\textbf{Algorithms / No. of Tasks} & \textbf{10}  & \textbf{12}  & \textbf{14}  & \textbf{16} & \textbf{18} & \textbf{20}  \\
\midrule
Random & 32.42	& 56.31	& 39	& 52.33	& 51.60	& 31.97  \\
Mix & 18.77	& 48.81	& 46.80	& 46.97	& 47.59	&17.21 \\
FIFO & 10.36	& 36.63	& 32.96	& 34.98	& 33.10	&6.14 \\
Greedy & 8.94	& 25.75	& 29.61	& 24.27	& 21.91	&4.09 \\
\bottomrule
\end{tabular*}
\end{table}
\end{center}

\begin{figure}
\centerline{\includegraphics[width=0.8\textwidth]{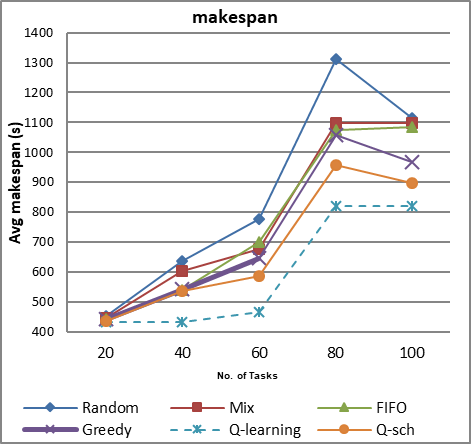}}
\caption{The average makespan in 2\textsuperscript{nd} scenario}\label{fig:makespan2}
\end{figure}

We have also conducted extended simulations to measure the average makespan of system in second scenario settings including more tasks and heterogeneous system setup. In this scenario, the length of tasks is smaller than task length in the firs scenario. Figure \ref{fig:makespan2} illustrates that the proposed algorithm performs well in this scenario. We can see from the results that, for example, the proposed algorithm achieves 37\%,25\%, 23\%, 22\% and 14\% improvement over random, mix, FIFO, greedy and Q-sch algorithms. The enhancement results for other experiments are represented in Table \ref{makespan2}.

\begin{center}
\begin{table}
\centering
\caption{Improving makespan of the proposed algorithm (2\textsuperscript{nd} scenario).\label{makespan2}}%
\begin{tabular*}{320pt}{@{\extracolsep\fill}lccccccc@{\extracolsep\fill}}
\toprule
\textbf{Algorithms / No. of Tasks}  & \textbf{20}  & \textbf{40}  & \textbf{60}  & \textbf{80} & \textbf{100}  \\
\midrule
Random & 4	&31.96	&39.87	&37.54	&26.30  \\
Mix & 3.13	&28.35	&31.16	&25.25	&25.15 \\
FIFO & 2.48	&19.85	&33.23	&23.57	&24.40 \\
Greedy & 2.04	&20.29	&27.86	&22.56	&15.01 \\
\bottomrule
\end{tabular*}
\end{table}
\end{center}

\begin{figure}
\centerline{\includegraphics[width=0.8\textwidth]{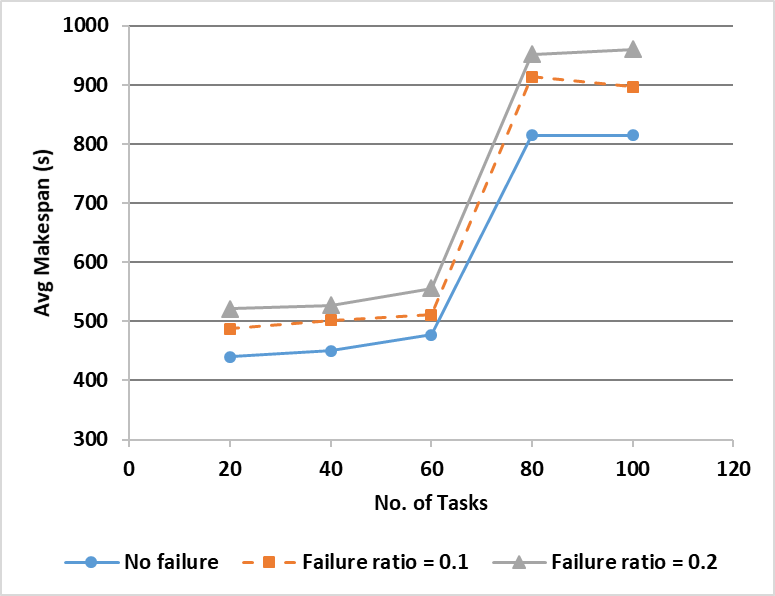}}
\caption{The effect of task failure on average makespan}\label{fig:failure}

\end {figure}

To evaluate the effect of task failure on the average makespan, we have designed a simulation scenario where an execution failure ratio has been set for each task, which means that each task has a small possibility that it would fail to complete. According to the our proposed algorithm, if a task failed to complete, it will be reprocessed in next scheduling rounds. The simulation has been executed in three cases: no failure, failure ratio = 0.1 and failure ratio = 0.2. As depicted in Figure \ref{fig:failure}, the average makespan for the submitted tasks increases 10 percent in average which indicates that the proposed QL-Scheduling algorithm is fairly robust to the scenarios with task failure.

\subsection{Average Response Time}
\begin{figure}
\centerline{\includegraphics[width=0.8\textwidth]{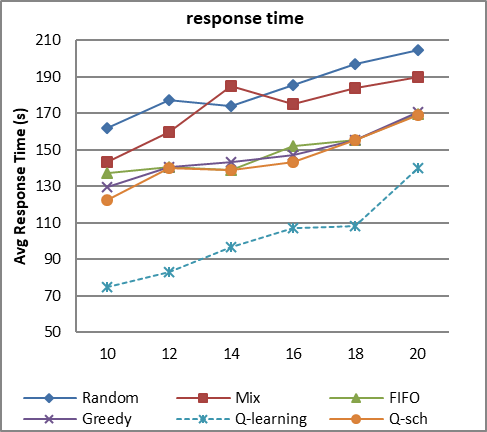}}
\caption{The average response time in first scenario}\label{fig:response1}
\end{figure}

The response time, as the total spent in the system to respond the users, is an important factor in the cloud performance evaluations. It is defined as the sum of task's service time and waiting time. The improvement in the average response time of cloud platforms substantially increases the users' satisfaction. Since in the proposed task scheduling algorithm, the average tasks' length in the virtual machines is considered in the learning process, the optimal policy results in shorter response time for submitted tasks. As identified in Figure (\ref{fig:response1}), the average response time is increased proportional to the tasks arrival rate. Due to the dynamic nature of task arrival to the system and unknown task length, the random, mix and FIFO algorithms does not function well in terms of response time. The greedy algorithm obtains similar results to FIFO approach. The states in the Q-sch algorithm are just defined based on the VM's empty buffer size and the system is rewarded based on the VM buffer space and waiting time which does not capture well the system dynamics. In the proposed method, with extended system state and reward function, shorter response time is attained compared to the existing algorithms (Table \ref{response1}). 

\begin{center}
\begin{table}
\centering
\caption{Improving average response time of the proposed algorithm (1\textsuperscript{st} scenario).\label{response1}}%
\begin{tabular*}{320pt}{@{\extracolsep\fill}lccccccc@{\extracolsep\fill}}
\toprule
\textbf{Algorithms / No. of Tasks} & \textbf{10}  & \textbf{12}  & \textbf{14}  & \textbf{16} & \textbf{18} & \textbf{20}  \\
\midrule
Random & 53.70	&53.10	&44.25	&42.47	&45.17	&31.70  \\
Mix & 47.55	&48.12	&47.56	&38.85	&41.30	&26.31 \\
FIFO & 45.25	&41.13	&30.21	&29.60	&30.32	&17.64 \\
Greedy & 42.30	&41.13	&32.16	&27.21	&30.32	&18.12 \\
\bottomrule
\end{tabular*}
\end{table}
\end{center}

\begin{figure}
\centerline{\includegraphics[width=0.8\textwidth]{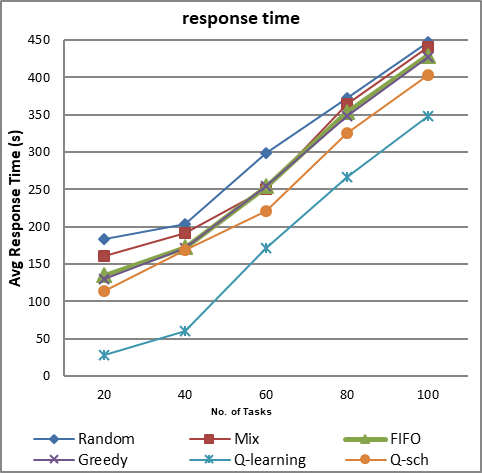}}
\caption{The average response time in second scenario}\label{fig:response2}
\end{figure}

In the second scenario, the proposed algorithm achieves better results in comparison to other algorithms. As illustrated in Figure (\ref{fig:response2}) the average response time has an increasing trend in terms of number of tasks. In the existing algorithms, a longer task may stall many shorter tasks in the VM queue. But, the optimal policy obtained from the Q-learning algorithm, intelligently distribute the tasks over the VMs with shorter queue size and therefore, improves remarkably the response time. The improvement in results for the proposed algorithm in the second scenario is shown in Table \ref{response2}.

\begin{center}
\begin{table}
\centering
\caption{Improving average response time of the proposed algorithm (2\textsuperscript{nd} scenario).\label{response2}}%
\begin{tabular*}{320pt}{@{\extracolsep\fill}lccccccc@{\extracolsep\fill}}
\toprule
\textbf{Algorithms / No. of Tasks}  & \textbf{20}  & \textbf{40}  & \textbf{60}  & \textbf{80} & \textbf{100}  \\
\midrule
Random & 84.69	&70.44	&42.28	&28.03	&22.19  \\
Mix &82.60	&68.58	&31.20	&26.64	&21.13 \\
FIFO & 79.10	&65.11	&32.28	&24.36	&19.11\\
Greedy & 78.46	&64.91	&32.28	&23.27	&18.73 \\
\bottomrule
\end{tabular*}
\end{table}
\end{center}

\subsection{Average Waiting Time}
\begin{figure}
\centerline{\includegraphics[width=0.8\textwidth]{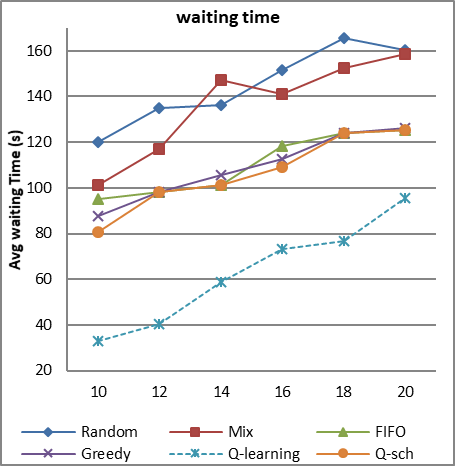}}
\caption{The average waiting time in first scenario}\label{fig:waiting1}
\end{figure}

\begin{figure}
\centerline{\includegraphics[width=0.8\textwidth]{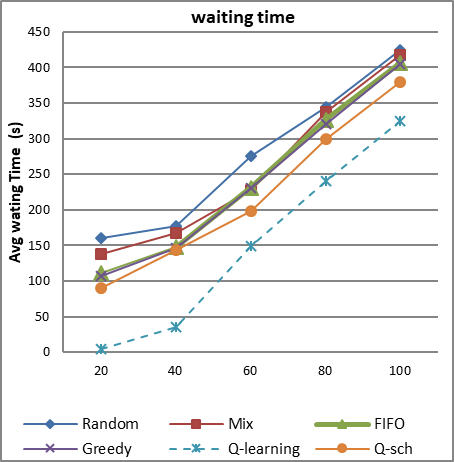}}
\caption{The average waiting time in second scenario}\label{fig:waiting2}
\end{figure}

The average waiting time includes the time interval where the tasks are queued in the VM, waiting to be executed over the cloud. Most task scheduling algorithms focus on improving this metric to satisfy the service-level agreements, boosting quality of service for the users. In this section, we conducted several experiments with two scenarios to assess this metric. As identified in Figure (\ref{fig:waiting1}), our proposed algorithm will result in shorter waiting times in comparison with the existing algorithms, since it does not hold the short length tasks behind the long tasks and assigns the tasks with the shortest queue length. The improvement of average waiting time in QL-Scheduling algorithm is maintained for different number of tasks (refer to Table \ref{waiting1}). In the second scenario with more task arrivals to the system (Figure (\ref{fig:waiting2})), the proposed algorithm outperforms the closest competitor, Q-sch algorithm, with 20\% improvement (Table \ref{waiting2}). In Q-sch algorithm, both waiting time and empty buffer size parameters are considered for the positive reward, while these parameters may conflict in some scenarios and penalize the algorithm performance. 

\begin{center}
\begin{table}
\centering
\caption{Improving average waiting time of the proposed algorithm (1\textsuperscript{st} scenario).\label{waiting1}}%
\begin{tabular*}{320pt}{@{\extracolsep\fill}lccccccc@{\extracolsep\fill}}
\toprule
\textbf{Algorithms / No. of Tasks} & \textbf{10}  & \textbf{12}  & \textbf{14}  & \textbf{16} & \textbf{18} & \textbf{20}  \\
\midrule
Random & 72.50	&70.37	&56.61	&51.65	&53.61	&40.62  \\
Mix & 67.32	&65.81	&59.86	&48.22	&49.67	&39.87 \\
FIFO & 65.26	&59.18	&41.58	&38.13	&37.90	&24 \\
Greedy & 62.50	&59.18	&44.33	&35.39	&37.90	&24.60 \\
\bottomrule
\end{tabular*}
\end{table}
\end{center}

\begin{center}

\begin{table}
\centering
\caption{Improving average waiting time of the proposed algorithm (2\textsuperscript{nd} scenario).\label{waiting2}}%
\begin{tabular*}{320pt}{@{\extracolsep\fill}lcccccc@{\extracolsep\fill}}
\toprule
\textbf{Algorithms / No. of Tasks} & \textbf{20}  & \textbf{40}  & \textbf{60}  & \textbf{80} & \textbf{100}  \\
\midrule
Random & 97.50	&80.22	&46.18	&30.52	&23.58  \\
Mix & 97.10	&78.91	&34.80	&29.08	&22.30 \\
FIFO & 96.39	&76.19	&35.93	&26.68	&20.19 \\
Greedy & 96.26	&76.02	&35.93	&25.13	&19.80 \\
\bottomrule
\end{tabular*}
\end{table}
\end{center}

\subsection{Average Utilization Ratio and Load Distribution}
\begin{figure}
\centerline{\includegraphics[width=0.8\textwidth]{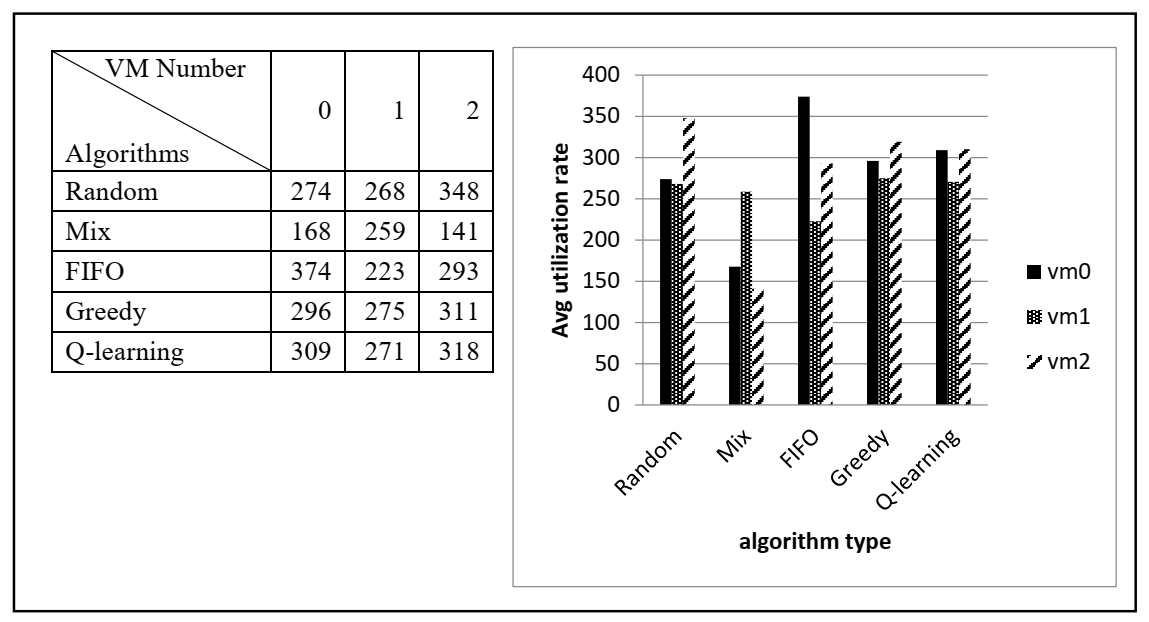}}
\caption{The average utilization in first scenario}\label{fig:avgutil}
\end{figure}

\begin{figure}
\centerline{\includegraphics[width=0.8\textwidth]{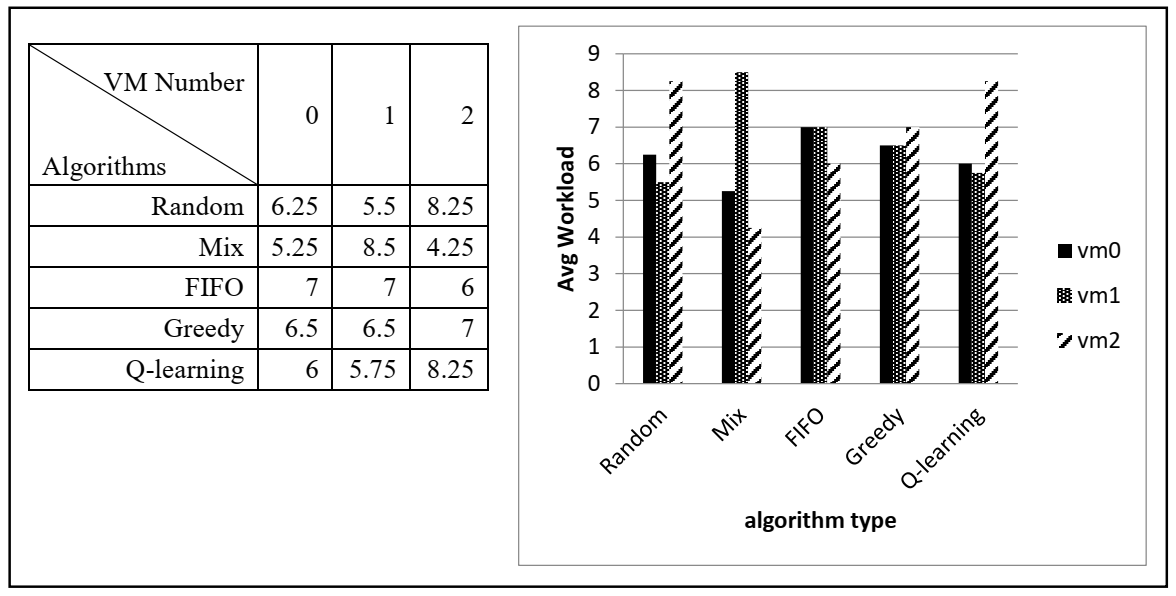}}
\caption{The average load distribution in first scenario}\label{fig:avgload}
\end{figure}

\begin{figure}
\centerline{\includegraphics[width=0.8\textwidth]{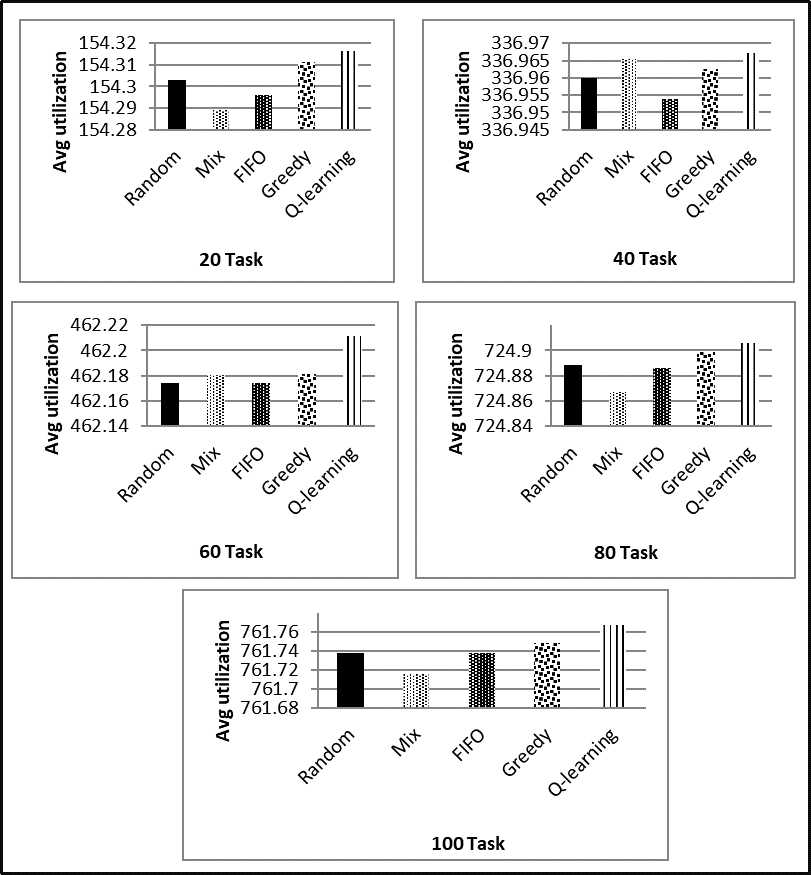}}
\caption{The average utilization in second scenario with 20-100 tasks}\label{fig:avgutilsecond}
\end{figure}

The task scheduling algorithms directly affects the load distribution over the virtual machines. From the perspective of system administrator, it is desirable to fully utilize the resources of virtual machines, while the load is evenly balanced over the VMs to reduce the power consumption of hosts. For evaluation of these metrics, simulations were performed for the first and second scenarios with different number of tasks. The results of average utilization ratio and average load distribution for three virtual machines in the first scenario are shown in Figure (\ref{fig:avgutil}) and Figure (\ref{fig:avgload}), respectively. The analysis of results reveals that the FIFO algorithm has poor performance in terms of load distribution since it assigns tasks based on the order of arrival. The greedy algorithm achieves better load distribution and improves the utilization ratio compared to FIFO method. The mix and random algorithms does not evenly distribute the system workload over the virtual machines. The proposed algorithm achieves good results for average utilization ratio of virtual machines, but can not evenly distribute tasks over the virtual machines. For further analysis, we have conducted extensive simulations on the second scenarios with more tasks (20 to 100). The results of simulations, illustrated in Figure (\ref{fig:avgutilsecond}), show that the proposed algorithm present better performance in terms of utilization ratio comparing to the other algorithms. Since the average length of tasks in this scenario is small, the difference between the average utilization of virtual machines for different algorithms will not be considerable. 


\subsection{The Effects of Buffer Size on Average Response Time}
\begin{figure}
\centerline{\includegraphics[width=0.8\textwidth]{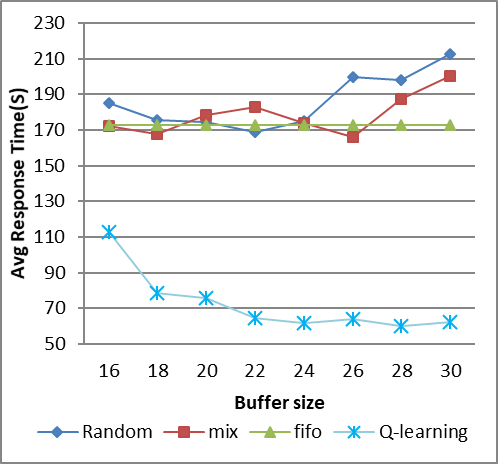}}
\caption{The effect of buffer size on average response time}\label{fig:buffer}
\end{figure}

In this section, the performance of the proposed algorithm and random, mix and FIFO algorithms with changes in buffer size has been assessed in a scenario with 40 tasks over 3 virtual machines with different buffer sizes. According to results of simulation in Figure (\ref{fig:buffer}), the average response time of mix and random algorithms increases with bigger buffer size. The FIFO algorithm does not change its performance with the changes in buffer size. The proposed algorithm shows much better performance in different buffer sizes compared to the other three algorithms. It is not sensitive to the changes in buffer size of virtual machines and results in a response time equals to 112 seconds, while the average response time for FIFO, mix and random algorithms equals to 183, 172 and 185 seconds, respectively. 

\section{Conclusion}
The optimal task allocation policy in the cloud computing platforms is a challenging work due to the dynamic and unpredictable nature of user tasks in the course of time. Most state-of-the-art solutions are myopic, meaning that they rely on the current state of the system and does not exploit the experiences from the previously-chosen actions over the time. In this paper, a stochastic approximation approach was applied to formulate the notion of learning for task allocation in a dynamic cloud environment. It was assumed that there is no \emph{a-priori} knowledge regarding time-varying nature of tasks in terms of task length and arrival time. More specifically, instead of proposing a rule-based or threshold-based approach, a model-free reinforcement learning solution was proposed which learns the optimal task allocation policy in the course of time. Our proposed algorithm is robust to high dynamics in the system due to its self-adaptive and self-learning capabilities. The exhaustive simulations revealed that the proposed algorithm outperforms the existing algorithms for task allocation. 

\section*{Acknowledgment}
We would like to express our gratitude to Fatemeh Ahmadi for her great assistance on implementing the simulation scenarios discussed in this article. 
\bibliographystyle{spphys}       
\bibliography{paperbib}   

\end{document}